\documentclass[aps,prd,reprint,preprintnumbers,showpacs,superscriptaddress,nofootinbib,floatfix]{revtex4-1}
\pdfoutput=1
\usepackage[colorlinks=true,citecolor=blue,linkcolor=blue]{hyperref}
\usepackage{amsmath,amssymb,bm,mathrsfs}
\usepackage{ulem}
\usepackage{graphicx}
\usepackage{bbding}
\usepackage{pifont}
\usepackage{booktabs}
\usepackage{longtable}
\usepackage{dcolumn}
\usepackage{ifsym}
\usepackage{slashed}
\usepackage[perpage,symbol*]{footmisc}
\begin{document}
\preprint{UCB-PTH-14/21,IPMU14-0102}

\title{Neutrino Mass Anarchy and the Universe}

\author{Xiaochuan Lu}
\email{luxiaochuan123456@berkeley.edu}
\affiliation{Department of Physics, University of California,
  Berkeley, California 94720, USA}
\affiliation{Theoretical Physics Group, Lawrence Berkeley National
  Laboratory, Berkeley, California 94720, USA}

\author{Hitoshi Murayama}
\email{hitoshi@berkeley.edu, hitoshi.murayama@ipmu.jp}
\affiliation{Department of Physics, University of California,
  Berkeley, California 94720, USA}
\affiliation{Theoretical Physics Group, Lawrence Berkeley National
  Laboratory, Berkeley, California 94720, USA}
\affiliation{Kavli Institute for the Physics and Mathematics of the
  Universe (WPI), Todai Institutes for Advanced Study, University of Tokyo,
  Kashiwa 277-8583, Japan}

\begin{abstract}
  We study the consequence of the neutrino mass anarchy on cosmology,
  in particular the total mass of neutrinos and baryon asymmetry
  through leptogenesis.  We require independence of measure in each
  mass matrix elements in addition to the basis independence, which
  uniquely picks the Gaussian measure.  A simple approximate $U(1)$
  flavor symmetry makes leptogenesis highly successful. Correlations
  between the baryon asymmetry and the light-neutrino quantities are
  investigated. We also discuss possible implications of recently
  suggested large total mass of neutrinos by the SDSS/BOSS data.
\end{abstract}

\maketitle

\section{Introduction \label{sec:Intro}}

Neutrino physics is a unique area in particle physics that has many
direct consequences on the evolution history and the current state of
Universe.  It was one of the first hypotheses for the non-baryonic
dark matter.  Excluding this possibility relied on rather surprising
constraint that the density of neutrinos would exceed that allowed by
Fermi degeneracy in the core of dwarf galaxies!
\cite{Tremaine:1979we} Because of the free streaming, massive neutrinos
would also suppress the large-scale structure, which is still subject
to active research.  The explosion mechanism of supernova is tied to
properties of neutrinos, and hence the chemical evolution of galaxies
depend on neutrinos.  The number of neutrinos is relevant to Big-Bang
Nucleosynthesis.  In addition, neutrinos may well have created the
baryon asymmetry of the Universe \cite{Fukugita:1986hr} or create the
Universe itself with scalar neutrino playing the role of the inflaton
\cite{Murayama:1993xu,Murayama:2014saa}.

Many of the consequences of neutrino properties on the Universe rely
on the mass of neutrinos.  After many decades of searches, neutrino
mass was discovered in 1998 in disappearance of atmospheric neutrinos
by the Super-Kamiokande collaboration \cite{Fukuda:1998mi}.
Subsequently the SNO experiment demonstrated transmutation of solar
electron neutrinos to other active neutrino species
\cite{Ahmad:2002jz} corroborated by reactor neutrino data from KamLAND
\cite{Eguchi:2002dm}.  Most recently, the last remaining mixing angle
was discovered by the Daya Bay reactor neutrino experiment
\cite{An:2012eh}.  Other experiments confirmed this discovery
\cite{Ahn:2012nd,Abe:2013hdq}.

On the other hand, fermion masses and mixings have been a great puzzle
in particle physics ever since the discovery of muon. Through decades
of intensive studies, we have discovered the existence of three
generations and a bizarre mass spectrum and mixings among them. The
underlying mechanism for this pattern is still not clear. But the
hierarchical masses and small mixings exhibited by quarks and charged
leptons seem to suggest that mass matrices are organized by some
yet-unknown symmetry principles. The discovery of neutrino masses and
mixings seem to even complicate the puzzle. From the current data
\cite{Beringer:1900zz}
\begin{eqnarray}
\Delta m_{12}^2 &=& (7.50 \pm 0.20) \times 10^{-5} \mbox{eV}^2, \\
\left| \Delta m_{23}^2 \right| &=& (2.32^{+0.12}_{-0.08}) \times 10^{-3} \mbox{eV}^2 ,\\
\sin^2 2\theta_{23} &>& 0.95 \hspace{0.2cm} (90\% \hspace{0.2cm} \text{C.L.}), \\
\sin^2 2\theta_{12} &=& 0.857 \pm 0.024, \\
\sin^2 2\theta_{13} &=& 0.095 \pm 0.010,
\end{eqnarray}
the neutrinos also display a small hierarchy $\Delta
m_{12}^2/\left|\Delta m_{23}^2\right| \sim \frac{1}{30}$
\footnote{Throughout this paper, we use {\it small hierarchy} for a
  mass spectrum typically like $1:3:10$, and {\it large hierarchy} for
  a mass spectrum typically like $1:10^2:10^4$. So in our wording, the
  neutrinos exhibit a small hierarchy, while the quarks and leptons
  exhibit a large hierarchy}, which is quite mild compared to quarks
and charged leptons. In addition, unlike quarks and charged leptons,
the neutrinos have both large and small mixing angles. Many attempts
were made to explain this picture using ordered, highly structured
neutrino mass matrices, especially when $\theta_{13}$ was consistent
with zero
\cite{Albright:1999ux,Altarelli:1998nx,Barbieri:1998jc,Blazek:1999ue,Nomura:1998gm}.

Quite counterintuitively, however, it was pointed out that the pattern
of neutrino masses and mixings can also be well accounted for by
structureless mass matrices \cite{Hall:1999sn}. Mass matrices with
independently random entries can naturally produce the small
hierarchical mass spectrum and the large mixing angles. This provides
us with an alternative point of view: instead of contrived models for
the mass matrix, one can simply take the random mass matrix as a low
energy effective theory \cite{Haba:2000be}. This anarchy approach is
actually more naturally expected, because after all, the three
generations possess the exact same gauge quantum numbers. From the
viewpoint of anarchy, however, the mass spectrum with large hierarchy
and small mixings exhibited by quarks and charged leptons need an
explanation. It turns out that introducing an approximate $U(1)$
flavor symmetry \cite{Froggatt:1978nt,Leurer:1992wg} can solve this
problem well \cite{Haba:2000be}. This combination of random mass
matrix and approximate $U(1)$ flavor symmetry has formed a new
anarchy-hierarchy approach to fermion masses and mixings
\cite{Haba:2000be}.

To be consistent with the spirit of anarchy, the measure to generate
the random mass matrices has to be basis independent
\cite{Haba:2000be}. This requires that the measure over the unitary
matrices be Haar measure, which unambiguously determines the
distributions of the mixing angles and CP phases. Consistency checks
of the predicted probability distributions of neutrino mixing angles
against the experimental data were also performed in great detail
\cite{deGouvea:2003xe,deGouvea:2012ac}. Although quite successful
already, this anarchy-hierarchy approach obviously has one missing
brick: a choice of measure to generate the eigenvalues of the random
mass matrices. With the only restriction being basis independence, one
can still choose any measure for the diagonal matrices at
will. However, as we will show in this paper, if in addition to basis
independence, one also wants the matrix elements to be distributed
independent from each other, then the only choice is the Gaussian
measure.

In this paper, we focus on the Gaussian measure to investigate the
consequences. As pointed out in \cite{Haba:2000be}, the quantities
most sensitive to this choice would be those closely related to
neutrino masses. We study a few such quantities of general interest,
including the effective mass of neutrinoless double beta decay
$m_{\text{eff}}$, the neutrino total mass $m_{\text{total}}$, and the
baryon asymmetry $\eta_{B0}$ obtained through a standard leptogenesis
\cite{Buchmuller:2002rq,Buchmuller:2004nz}. We also present a
correlation analysis between $\eta_{B0}$ and light-neutrino
parameters. Recently, the correlation between leptogenesis and
light-neutrino quantities was also studied by taking linear measure in
\cite{Jeong:2012zj}. Their results are in broad qualitative agreement
with ours in this paper.

The rest of this paper is organized as following. We first motivate our sampling model---Gaussian measure combined with approximate U(1) flavor symmetry---in Section~\ref{sec:Sampling}. Then the consequences of this sampling model is presented in Section~\ref{sec:Consequences}. We show our Monte Carlo predictions on light-neutrino mass hierarchy, effective mass of neutrinoless double beta decay, light-neutrino total mass, and baryon asymmetry through leptogenesis. In Section~\ref{sec:Correlation}, we investigate the correlations between baryon asymmetry and light-neutrino quantities. A recent Baryon Oscillation Spectroscopic Survey (BOSS) analysis suggests that the total neutrino mass could be quite large \cite{Beutler:2014yhv}. Our Section~\ref{sec:MtotalCut} is devoted to discuss the consequence of this suggestion. We conclude in Section~\ref{sec:Conclusion}.

\section{Sampling Model \label{sec:Sampling}}

\subsection{Gaussian Measure}

To accommodate the neutrino masses, we consider the standard model with an addition of three generations of right-handed neutrinos $\nu_R$, which are singlets under $SU(2)_L \times U(1)_Y$ gauge transformations. Then there are two neutrino mass matrices, the {\it Majorana mass matrix} $m_R$ and the {\it Dirac mass matrix} $m_D$, allowed by gauge invariance
\begin{eqnarray}
 \Delta {\cal L} &\supset& -\epsilon^{ab} {\bar L}_a H_b^\dag y_\nu \nu_R -\frac{1}{2} {\bar\nu}_R^c m_R \nu_R + h.c. \nonumber \\
 &\supset& -{\bar\nu}_L {m_D} \nu_R -\frac{1}{2} {\bar\nu}_R^c m_R \nu_R + h.c., \label{eq:Lagrangian}
\end{eqnarray}
where $y_\nu=\frac{\sqrt 2}{v}m_D$, with $v=246$~GeV. With the spirit of anarchy, we should not discard either of them without any special reason. Both should be considered as random inputs. We parameterize them as
\begin{eqnarray}
 m_R &=& {\cal M} \cdot U_R D_R U_R^T, \label{eq:paramR}\\
 m_D &=& {\cal D} \cdot U_1 D_0 U_2^\dag, \label{eq:paramD}
\end{eqnarray}
where $D_R$ and $D_0$ are real diagonal matrices with non-negative elements; $U_R$, $U_1$ and $U_2$ are unitary matrices; ${\cal M}$ and ${\cal D}$ are overall scales.

At this point, the requirement of neutrino basis independence turns out to be very powerful. It requires that the whole measure of the mass matrix factorizes into that of the unitary matrices and diagonal matrices \cite{Haba:2000be}:
\begin{eqnarray}
 dm_R &\sim& dU_R dD_R, \label{eq:decom1} \\
 dm_D &\sim& dU_1 dU_2 dD_0. \label{eq:decom2}
\end{eqnarray}
Furthermore, it also demands the measure of $U_R$, $U_1$ and $U_2$ to be the Haar measure of $U(3)$ group \cite{Haba:2000be}:
\begin{eqnarray}
U &=& {e^{i\eta }}{e^{i{\phi _1}{\lambda _3} + i{\phi _2}{\lambda _8}}}\left( {\begin{array}{*{20}{c}}
   1 & 0 & 0  \\
   0 & {{c_{23}}} & {{s_{23}}}  \\
   0 & { - {s_{23}}} & {{c_{23}}}  \\
\end{array}} \right) \nonumber \\
 && \times \left( {\begin{array}{*{20}{c}}
   {{c_{13}}} & 0 & {{s_{13}}{e^{ - i\delta }}}  \\
   0 & 1 & 0  \\
   { - {s_{13}}{e^{i\delta }}} & 0 & {{c_{13}}}  \\
\end{array}} \right) \nonumber \\
 && \times \left( {\begin{array}{*{20}{c}}
   {{c_{12}}} & {{s_{12}}} & 0  \\
   { - {s_{12}}} & {{c_{12}}} & 0  \\
   0 & 0 & 1  \\
\end{array}} \right){e^{i{\chi _1}{\lambda _3} + i{\chi _2}{\lambda _8}}}, \\
dU &=& ds_{23}^2dc_{13}^4ds_{12}^2d\delta  \cdot d{\chi _1}d{\chi _2} \cdot d\eta d{\phi _1}d{\phi _2},
\end{eqnarray}
where $\lambda_3=\text{diag}(1,-1,0)$, $\lambda_8=\text{diag}(1,1,-2)/\sqrt{3}$, and $c_{12}=\cos\theta_{12}$, etc.

Although basis independence is very constraining, it does not uniquely fix the measure choice of $m_R$ or $m_D$, because arbitrary measure of the diagonal matrices $D_R$ and $D_0$ is still allowed. Now let us look at the entries of $m_R$ and $m_D$. There are 9 complex free entries for $m_D$ and 6 complex free entries for the symmetric matrix $m_R=m_R^T$. Generating each free entry independently is probably the most intuitive way of getting a random matrix. However, if one combines this free entry independence with the basis independence requirement, then it turns out there is only one option left---the Gaussian measure:
\begin{eqnarray}
 d{m_D} &\sim& \prod\limits_{ij} e^{-\left|m_{D,ij}\right|^2} dm_{D,ij} \nonumber \\
 &=& \left( \prod\limits_{ij} dm_{D,ij} \right) e^{-tr(m_D m_D^\dagger)}\ , \label{eq:mDgaussian} \\
 d{m_R} &\sim& \prod\limits_i e^{-\left|m_{R,ii}\right|^2} dm_{R,ii} \prod\limits_{i<j} e^{-2\left|m_{R,ij}\right|^2} dm_{R,ij} \nonumber \\
 &=& \left( \prod\limits_{i \le j} dm_{R,ij} \right) e^{-tr(m_R m_R^\dagger)}\ . \label{eq:mRgaussian}
\end{eqnarray}
Although this is a known result in random matrix theory \cite{mehta2004random,Bai:2012zn}, we also present our own proof in the appendix.

On one hand, basis independence is necessary from the spirit of anarchy. On the other hand, free entry independence is also intuitively preferred. With these two conditions combined, we are led uniquely to the Gaussian measure. Now the only parameters left free to choose are the two units ${\cal M}$ and ${\cal D}$. Following the spirit of anarchy, ${\cal D}$ should be chosen to make the neutrino Yukawa coupling of order unity,
\begin{equation}
y_\nu=\frac{\sqrt 2}{v} m_D \sim O(1) .
\end{equation}
Throughout this paper, we choose ${\cal D}=30$~GeV, which yields $y_\nu \sim 0.6$. Then we choose ${\cal M}$ to fix the next largest mass-square difference of light-neutrinos $\Delta m_l^2$ at $2.5 \times 10^{-3}$~eV$^2$ in accordance with the data.

\subsection{Approximate $U(1)$ Flavor Symmetry}

Our model (Eq.(\ref{eq:Lagrangian})) is capable of generating a baryon asymmetry $\eta_{B0}$ through thermal leptogenesis \cite{Fukugita:1986hr,Luty:1992un}. For the simplicity of analysis, we would like to focus on the scenario with two conditions:

(1) There is a hierarchy among heavy-neutrino masses $M_1 \ll M_2, M_3$, so that the dominant contribution to $\eta_{B0}$ is given by the decay of the lightest heavy neutrino $N_1$ \cite{Buchmuller:2002rq}.

(2) If we use $h_{ij}$ to denote the Yukawa couplings of heavy-neutrino mass eigenstates
\begin{equation}
\Delta{\cal L} \supset -h_{ij} \epsilon^{ab} {\bar L}_{ai} H_b^\dag N_j,
\end{equation}
then the condition $h_{i1} \ll 1$ for all $i=1,2,3$ would justify the neglect of annihilation process $N_1 N_1 \to l {\bar l}$, and also help driving the decay of $N_1$ out of equilibrium \cite{Luty:1992un,Buchmuller:2002rq}. This condition used to be taken as an assumption \cite{Luty:1992un}.

\renewcommand\arraystretch{1.4}
\begin{table}
\centering
\begin{tabular}{|c|c|}
  \hline
  $\hspace{0.2cm}$ right-handed neutrino $\hspace{0.2cm}$ & $\hspace{0.2cm}$ $U(1)$ flavor charge $\hspace{0.2cm}$ \\
  \hline
  ${\nu _{R,1}}$ & 2 \\
  ${\nu _{R,2}}$ & 1 \\
  ${\nu _{R,3}}$ & 0 \\
  \hline
\end{tabular}
\caption{\label{tbl:U1charge} The $U(1)$ flavor charge assignments for right-handed neutrinos used in this paper.}
\end{table}
\renewcommand\arraystretch{1}

Both conditions above can be achieved by imposing a simple $U(1)$ flavor charge on the right-handed neutrinos. For example, one can make charge assignments as shown in Table.~\ref{tbl:U1charge}. Assuming a scalar field $\phi$ carries $-1$ of this $U(1)$ flavor charge, one can construct neutral combinations $\nu _{\phi}$ as
\begin{equation}
\nu_\phi=\left( {\begin{array}{*{20}{c}}
   {{\nu _{\phi ,1}}}  \\
   {{\nu _{\phi ,2}}}  \\
   {{\nu _{\phi ,3}}}  \\
\end{array}} \right) = \left( {\begin{array}{*{20}{c}}
   {{\nu _{R,1}} \cdot {\phi ^2}}  \\
   {{\nu _{R,2}} \cdot \phi }  \\
   {{\nu _{R,3}} \cdot 1}  \\
\end{array}} \right) .
\end{equation}
Now it only makes sense to place the random coupling matrices among these neutral combinations
\begin{eqnarray}
 \Delta{\cal L} &\supset& -{\bar\nu}_L m_{D0} {\nu_\phi} -\frac{1}{2}{\bar\nu}_\phi^c m_{R0} {\nu_\phi} + h.c., \label{eq:newLagrangian}
\end{eqnarray}
where $m_{R0}$ and $m_{D0}$ should be generated according to Gaussian measure as in Eq.~(\ref{eq:mDgaussian})-(\ref{eq:mRgaussian}). Upon $U(1)$ flavor symmetry breaking $\left\langle \phi \right\rangle=\epsilon$ with $\epsilon \simeq 0.1$, this gives
\begin{equation}
{\nu _\phi } \supset \left( {\begin{array}{*{20}{c}}
   {{\epsilon ^2}} & 0 & 0  \\
   0 & \epsilon  & 0  \\
   0 & 0 & 1  \\
\end{array}} \right) \cdot {\nu _R} \equiv {D_\epsilon } \cdot {\nu _R},
\end{equation}
and hence the mass matrices
\begin{eqnarray}
 {m_R} &=& {D_\epsilon }{m_{R0}}{D_\epsilon } = {\cal M} \cdot {D_\epsilon }{U_R}{D_R}{U_R}^T{D_\epsilon } , \\
 {m_D} &=& {m_{D0}}{D_\epsilon } = {\cal D} \cdot {U_1}{D_0}{U_2}^\dag {D_\epsilon }.
\end{eqnarray}
Let us parameterize the heavy-neutrino mass matrix as $m_N = U_N M U_N^T$, then
\begin{eqnarray}
m_N &\approx& m_R \sim \left( {\begin{array}{*{20}{c}}
   {{\epsilon ^4}} & {{\epsilon ^3}} & {{\epsilon ^2}} \\
   {{\epsilon ^3}} & {{\epsilon ^2}} & \epsilon   \\
   {{\epsilon ^2}} & \epsilon  & 1  \\
\end{array}} \right) \nonumber \\
&\sim& \left( {\begin{array}{*{20}{c}}
   1 & \epsilon  & {{\epsilon ^2}}  \\
   \epsilon  & 1 & \epsilon   \\
   {{\epsilon ^2}} & \epsilon  & 1  \\
\end{array}} \right)\left( {\begin{array}{*{20}{c}}
   {{\epsilon ^4}} & 0 & 0  \\
   0 & {{\epsilon ^2}} & 0  \\
   0 & 0 & 1  \\
\end{array}} \right)\left( {\begin{array}{*{20}{c}}
   1 & \epsilon  & {{\epsilon ^2}}  \\
   \epsilon  & 1 & \epsilon   \\
   {{\epsilon ^2}} & \epsilon  & 1  \\
\end{array}} \right) ,
\end{eqnarray}
where one can identify
\begin{equation}
M \sim \left( {\begin{array}{*{20}{c}}
   {{\epsilon ^4}} & 0 & 0  \\
   0 & {{\epsilon ^2}} & 0  \\
   0 & 0 & 1  \\
\end{array}} \right)\ , \mbox{ } \mbox{ } \mbox{ }
{U_N} \sim \left( {\begin{array}{*{20}{c}}
   1 & \epsilon  & {{\epsilon ^2}}  \\
   \epsilon  & 1 & \epsilon   \\
   {{\epsilon ^2}} & \epsilon  & 1  \\
\end{array}} \right) .
\end{equation}
Clearly a hierarchy among heavy neutrino masses is achieved. Furthermore, the heavy neutrino mass eigenstates are $N=U_N^T\nu_R$. Since
\begin{equation}
\Delta{\cal L} \supset -\epsilon^{ab} {\bar L}_a H_b^\dag y_\nu \nu_R \equiv -h_{ij} \epsilon^{ab} {\bar L}_{ai} H_b^\dag N_j,
\end{equation}
we obtain the Yukawa coupling $h_{ij}$ as
\begin{eqnarray}
h &=& y_\nu U_N^* = \frac{\sqrt 2}{v} m_D U_N^* \nonumber \\
&\sim& \frac{{\sqrt 2 }}{v}{m_{D0}}\left( {\begin{array}{*{20}{c}}
   {{\epsilon ^2}} & 0 & 0  \\
   0 & \epsilon  & 0  \\
   0 & 0 & 1  \\
\end{array}} \right)\left( {\begin{array}{*{20}{c}}
   1 & \epsilon  & {{\epsilon ^2}}  \\
   \epsilon  & 1 & \epsilon   \\
   {{\epsilon ^2}} & \epsilon  & 1  \\
\end{array}} \right) \nonumber \\
&\sim& \frac{{\sqrt 2 }}{v}{m_{D0}}\left( {\begin{array}{*{20}{c}}
   {{\epsilon ^2}} & {{\epsilon ^3}} & {{\epsilon ^4}}  \\
   {{\epsilon ^2}} & \epsilon  & {{\epsilon ^2}}  \\
   {{\epsilon ^2}} & \epsilon  & 1  \\
\end{array}} \right) .
\end{eqnarray}
We see that ${h_{i1}} \sim {\epsilon ^2} \ll 1$ is guaranteed for all $i=1,2,3$.

It is worth noting that the light-neutrino mass matrix $m_\nu$ is not
affected by our U(1) flavor change assignment on right-handed
neutrinos (Table.~\ref{tbl:U1charge}). The hierarchical matrix
$D_\epsilon$ cancels due to the seesaw mechanism
\cite{Minkowski:1977sc,Yanagida:1979as,GellMann:1980vs,Yanagida:1980xy,Glashow:1979nm,Mohapatra:1979ia}:
\begin{equation}
m_\nu = m_D m_R^{-1} m_D^T = m_{D0} m_{R0}^{-1} m_{D0}^T.
\end{equation}
Therefore all properties of light neutrinos do not depend on the
particular U(1) flavor charge assignment nor the size of the breaking
parameter $\epsilon$.  The leptogenesis aspect is the only discussion
in the paper where this flavor structure is relevant.

\section{Consequences \label{sec:Consequences}}

In this section, we present our Monte Carlo results based on the
sampling measure described in the last section.

\subsection{Light-neutrino Mixings and Mass Splitting Ratio}

We parameterize the light neutrino mass matrix as
\begin{equation}
m_\nu = m_D m_R^{-1} m_D^T \equiv U_\nu m U_\nu^T,
\end{equation}
with $U_\nu$ a unitary matrix and $m$ a real diagonal matrix with
non-negative elements. We also follow a conventional way to label the
three masses of the light neutrinos: First sort them in the ascending
order $m_{11} \le m_{22} \le m_{33}$. Then there are two mass-squared
splittings $m_{22}^2 - m_{11}^2$ and $m_{33}^2 - m_{22}^2$. We use
$\Delta m_s^2$ and $\Delta m_l^2$ to denote the smaller and larger one
respectively. If $\Delta m_s^2$ is the first one, we call this
scenario ``normal" and take the definitions $m_1 \equiv m_{11}, m_2
\equiv m_{22}, m_3 \equiv m_{33}$. Otherwise, we call it ``inverted"
and take $m_1 \equiv m_{22}, m_2 \equiv m_{33}, m_3 \equiv
m_{11}$. The columns of the unitary matrix $U_\nu$ should be arranged
accordingly.

Predictions on light-neutrino mixings---the distributions of mixing
angles $\theta_{12},\theta_{23},\theta_{13}$, CP phase $\delta_{CP}$,
and other physical phases $\chi_1,\chi_2$---are certainly the same as
in general study of basis independent measures \cite{Haba:2000be},
since $U_\nu$ is totally governed by the Haar measure. A statistical
Kolmogorov-Smirnov test shows that the predicted mixing angle
distribution is completely consistent with the experimental
data\cite{deGouvea:2003xe,deGouvea:2012ac}.

\begin{figure}[tpb]
 \centering
 \includegraphics[width=6cm]{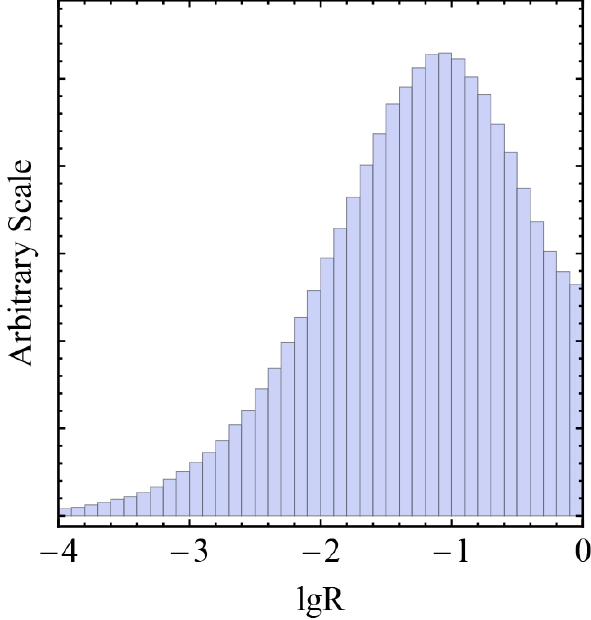}
 \caption{Histogram of $\text{lg}\ R = \text{log}_{10} R$ with $10^6$
   occurrences collected. \label{fig:lgR}}
\end{figure}

The mass-squared splitting ratio $R \equiv \Delta m_s^2/\Delta m_l^2$ is observed to be around \cite{Beringer:1900zz}
\begin{equation}
R_{\text{exp}} \equiv \frac{7.50 \times 10^{-5}}{2.32 \times 10^{-3}}.
\end{equation}
Fig.~\ref{fig:lgR} shows our predicted distribution of this ratio. With a probability of $R < R_{\text{exp}}$ being $36.1\%$, the prediction is completely consistent with the data.

For the purpose of studying other quantities, we introduce the
following {\it Mixing-Split cuts} as suggested by the experimental
data \cite{Beringer:1900zz} on light-neutrino mixings and mass-squared
splitting ratio:
\begin{eqnarray}
&& {\sin^2}2{\theta_{23}} = 1.0 \label{eq:cut23} \\
&& {\sin^2}2{\theta_{12}} = 0.857 \label{eq:cut12} \\
&& {\sin^2}2{\theta_{13}} = 0.095 \label{eq:cut13} \\
&& R \in R_{\text{exp}} \times (1-0.05, 1+0.05) \label{eq:cutR}
\end{eqnarray}

\subsection{Mass Hierarchy, $m_{\text{eff}}$ and $m_{\text{total}}$}

\begin{figure}[tpb]
 \centering
 \includegraphics[width=6cm]{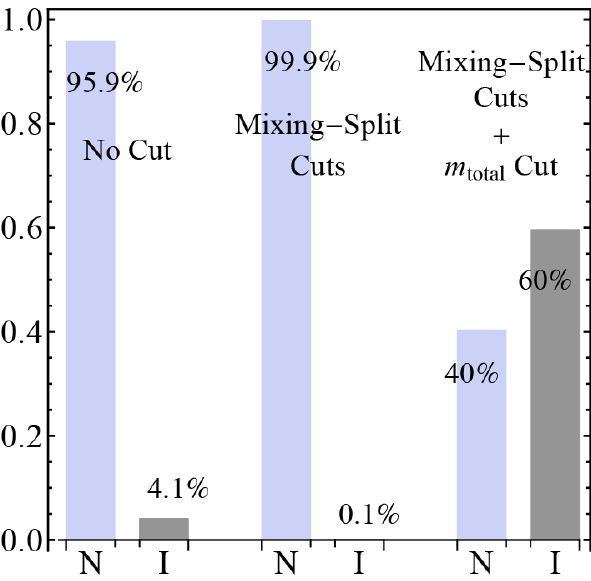}
 \caption{Fractions of normal and inverted mass hierarchy scenarios under different cuts selections, where ``N" stands for normal hierarchy and ``I" stands for inverted hierarchy. Each of the first two columns consists of $10^6$ occurrences, while the last column ``Mixing-Split Cuts + $m_\text{total}$ Cut" contains only $10^4$ occurrences. \label{fig:hierarchy}}
\end{figure}

For the mass hierarchy of light-neutrino, our anarchy model predicts normal scenario being dominant. A $10^6$ sample Monte Carlo finds the fraction of normal scenario being $95.9\%$ before the Mixing-Split cuts (Eq.~(\ref{eq:cut23})-(\ref{eq:cutR})), and $99.9\%$ after applying the cuts. Fig.~\ref{fig:hierarchy} shows the fractions of each scenario.

\begin{figure*}[tpb]
 \centering
 \includegraphics[width=13cm]{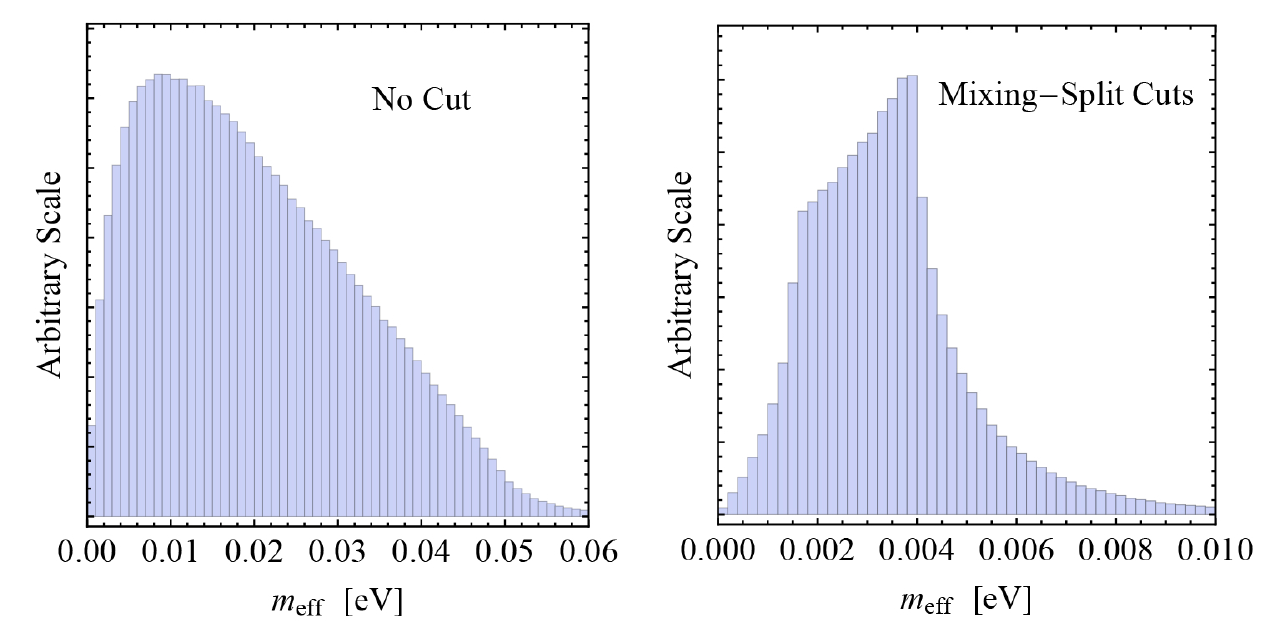}\nonumber
 \caption{Histogram of $m_\text{eff}$ with $10^6$ occurrences collected. Left/Right panel shows distribution before/after applying the Mixing-Split cuts. \label{fig:meff}}
\end{figure*}

\begin{figure*}[tbp]
 \centering
 \includegraphics[width=13cm]{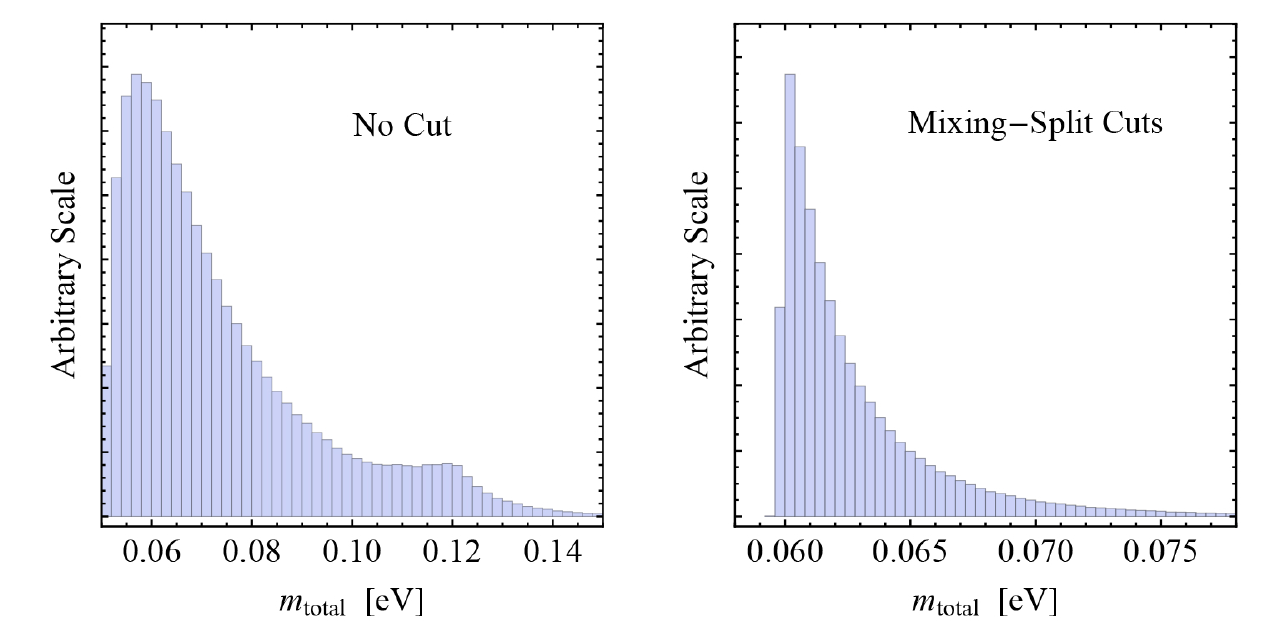}
 \caption{Histogram of $m_\text{total}$, with $10^6$ occurrences collected. Left/Right panel shows distribution before/after applying the Mixing-Split cuts. \label{fig:mtotal}}
\end{figure*}

Neutrino anarchy allows for a random, nonzero Majorana mass matrix $m_R$. This means that the light neutrinos are Majorana and thus there can be neutrinoless double beta decay process. The effective mass of this process $m_{\text{eff}} \equiv \left| \sum\limits_i m_i{U_{v,ei}^2} \right|$ is definitely a very broadly interested quantity. Another quantity of general interest is the light-neutrino total mass $m_\text{total} \equiv m_1 + m_2 + m_3$, due to its presence in cosmological processes. Our predictions on $m_{\text{eff}}$ and $m_{\text{total}}$ are shown in Fig.~\ref{fig:meff} and Fig.~\ref{fig:mtotal} respectively. For each quantity, we plot both its whole distribution under Gaussian measure and its distribution after applying the Mixing-Split cuts. Distributions of $m_\text{eff}$ and $m_\text{total}$ under Gaussian measure were also studied recently in \cite{Bergstrom:2014owa}. Their results are in agreement with ours. The small difference is due to the difference in taking cuts on neutrino mixing angles.

\begin{figure*}[tbp]
 \centering
 \includegraphics[width=13cm]{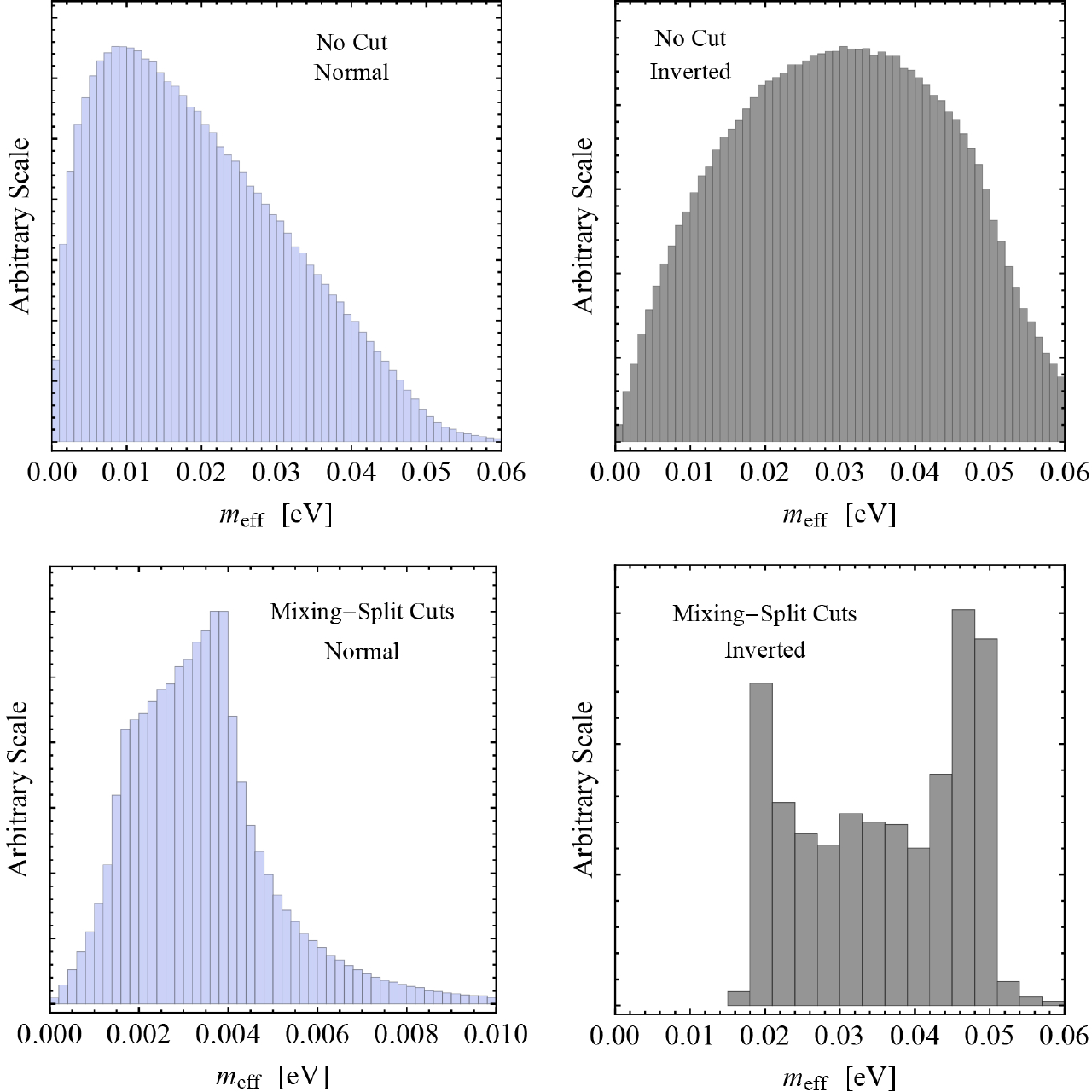}
 \caption{Histogram of $m_\text{eff}$ with two mass hierarchy scenarios plotted separately. Left/Right column shows distribution under normal/inverted scenario. Upper/Lower row shows distribution before/after applying the Mixing-Split cuts. The plot of inverted scenario with Mixing-Split cuts applied (right bottom) contains $10^4$ occurrences, while other plots contain $10^6$ occurrences. \label{fig:meffNI}}
\end{figure*}

\begin{figure*}[tbp]
 \centering
 \includegraphics[width=13cm]{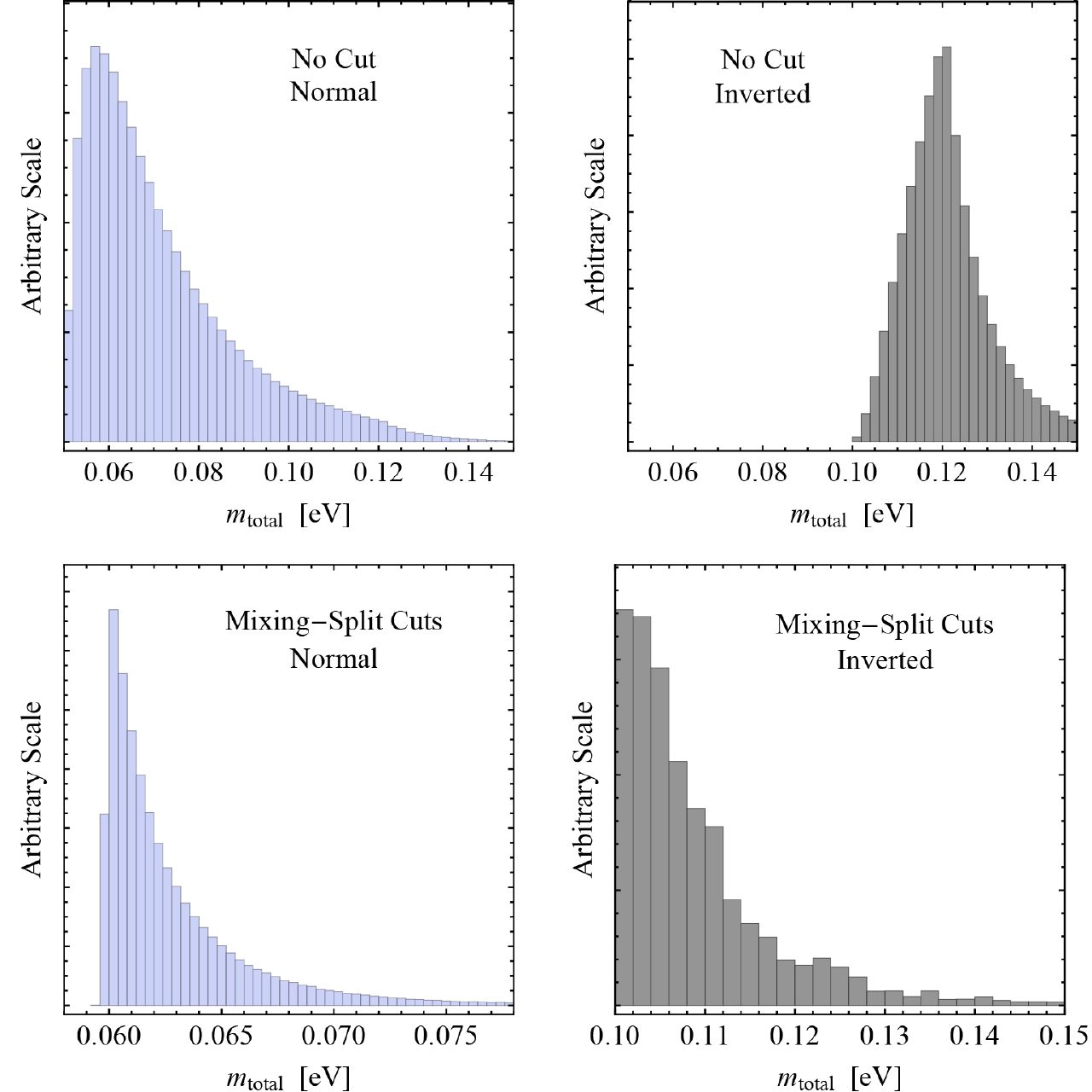}
 \caption{Histogram of $m_\text{total}$ with two mass hierarchy scenarios plotted separately. Left/Right column shows distribution under normal/inverted scenario. Upper/Lower row shows distribution before/after applying the Mixing-Split cuts. The plot of inverted scenario with Mixing-Split cuts applied (right bottom) contains $10^4$ occurrences, while other plots contain $10^6$ occurrences. \label{fig:mtotalNI}}
\end{figure*}

We see from Fig.~\ref{fig:meff} that most of the time $m_\text{eff}
\lesssim 0.05$~eV. It becomes even smaller after we apply the
Mixing-Split Cuts, mainly below $m_\text{eff} \lesssim 0.01$~eV. This
is very challenging to experimental sensitivity. For $m_\text{total}$,
Fig.~\ref{fig:mtotal} shows it being predicted to be very close to the
current lower bound $\sim 0.06$~eV. The kink near $0.1~$eV is due to
the superposition of the two mass hierarchy scenarios.

To understand each component better, we show the distributions of
$m_\text{eff}$ and $m_\text{total}$ in Fig.~\ref{fig:meffNI} and
Fig.~\ref{fig:mtotalNI} for both before/after applying the cuts and
normal/inverted hierarchy scenario. As Fig.~\ref{fig:mtotalNI} shows,
under either hierarchy scenario, $m_\text{total}$ is likely to reside
very close to its corresponding lower bound, especially after applying
the cuts. Very interestingly, however, recent BOSS analysis suggests
$m_\text{total}$ could be quite large, with a center value $\sim
0.36$~eV \cite{Beutler:2014yhv}. As seen clearly from
Fig.~\ref{fig:mtotalNI}, a large value of $m_\text{total}$ would
favor inverted scenario. We will discuss some possible consequences
of this suggestion in Section~\ref{sec:MtotalCut}.

\subsection{Leptogenesis}

As explained in Section~\ref{sec:Sampling}, with our approximate $U(1)$ flavor symmetry, the baryon asymmetry $\eta_{B0} \equiv \frac{n_B}{n_\gamma}$ can be computed through the standard leptogenesis calculations \cite{Buchmuller:2002rq,Buchmuller:2004nz}. For each event of $m_R$ and $m_D$ generated, we solve the following Boltzmann equations numerically.
\begin{eqnarray}
 \frac{dN_1}{dz} &=& -(N_1-N_1^\text{eq})(D+S) ,\label{eq:diff1} \\
 \frac{dN_{B-L}}{dz} &=& -(N_1-N_1^\text{eq}) \varepsilon D - N_{B-L} W, \label{eq:diff2}
\end{eqnarray}
where we have followed the notations in \cite{Buchmuller:2002rq} and \cite{Buchmuller:2004nz}.

\begin{figure*}[tpb]
 \centering
 \includegraphics[width=13cm]{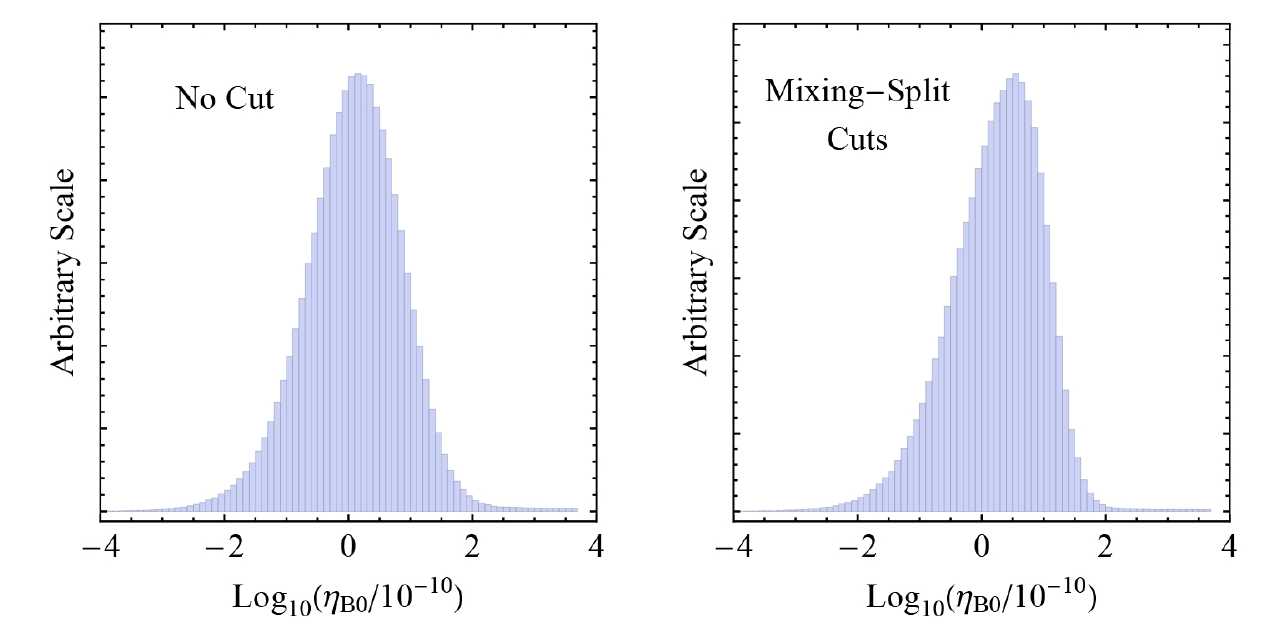}
 \caption{Histogram of $\eta_{B0}$ with $10^6$ occurrences collected. Left/Right panel shows distribution before/after applying Mixing-Split cuts. \label{fig:LogetaB0}}
\end{figure*}

The argument $z \equiv M_1/T$ is evolved from $z_i=0.001$ to
$z_f=20.0$. Evolving $z$ further beyond $20.0$ is not necessary,
because the value of $N_{B-L}$ becomes frozen shortly after
$z>10.0$. The baryon asymmetry is then given by $\eta_{B0}=0.013
N_{B-L}^0 \approx 0.013 N_{B-L}(z_f)$ \cite{Buchmuller:2002rq}. Due to
randomly generated $m_R$ and $m_D$, we have equal chances for
$\varepsilon$ to be positive or negative. It is the absolute value
that is meaningful. We take the initial condition $N_{B-L}(z_i)=0$.
We actually tried two typical initial conditions for $N_1$,
$N_1(z_i)=0$ and $N_1(z_i)=N_1^\text{eq}(z_i)$, and found no
recognizable differences. The distributions of $\eta_{B0}$, both
before and after applying the Mixing-Split cuts, are shown in
Fig.~\ref{fig:LogetaB0}. We see from figure that our prediction on
$\eta_{B0}$ is about the correct order of magnitude as
$\eta_{B0,\text{exp}} \approx 6\times 10^{-10}$ \cite{Komatsu:2010fb}.

Let us try to understand the results from some crude estimates. First, let us see why there is almost no difference between the two initial conditions $N_1(z_i)=0$ and $N_1(z_i)=N_1^\text{eq}(z_i)$. The decay function $D(z)$ has the form \cite{Buchmuller:2004nz}:
\begin{equation}
D(z) = {\alpha_D}\frac{{{K_1}(z)}}{{{K_2}(z)}}z,
\end{equation}
where $K_1(z)$ and $K_2(z)$ are modified Bessel functions, and the constant $\alpha_D$ is proportional to the {\it effective neutrino mass} ${\tilde m}_1$:
\begin{equation}
\alpha_D \propto {\tilde m}_1 \equiv \frac{(m_D^\dag m_D)_{11}}{M_1}\ .
\end{equation}
So roughly speaking, after $z \gtrsim 1$, the modified Bessel
functions become rather close and $D(z)$ increases linearly with
$z$. But in our prescription, the value of $D(z)+S(z)$ becomes already
quite large, typically around $50$, at $z=1.0$. So $N_1$ is forced to
be very close to $N_1^{\text{eq}}$ thereafter and the solution to the
first differential equation (Eq.~(\ref{eq:diff1})) is approximately
\begin{equation}
N_1-N_1^{\text{eq}} = -\frac{1}{D+S}\frac{dN_1}{dz} \approx
-\frac{1}{D+S} \frac{dN_1^\text{eq}}{dz}\ .
\end{equation}
Of course, this initial-condition-independent solution only holds when $D+S$ is large enough, typically for $z>1.0$. The values of $N_1-N_1^\text{eq}$ at $z<1.0$ certainly have a considerable dependence on $N_1(z_i)$. However, the solution to the second differential equation (Eq.(\ref{eq:diff2})) is
\begin{equation}
  N_{B-L}^0 = \varepsilon \cdot \int_0^\infty  {dz\frac{D}{{D +
        S}}\frac{{dN_1}}{{dz}}{e^{ - \int_z^\infty  {W(z')dz'} }}} \ .
  \label{eq:efficiency}
\end{equation}
And due to the shape of $W(z)$, yield of $N_{B - L}^0$ at $z<1.0$ is significantly suppressed by a factor $e^{-\int_z^\infty {W(z')dz'}}$. Therefore $\eta_{B0}$ turns out to be almost independent of $N_1(z_i)$.

Second, let us crudely estimate the order of magnitude of $\eta_{B0}$. In addition to $D(z)$, the scattering functions $S(z)$ and washout function $W(z)$ are also proportional to ${\tilde m}_1$. So ${\tilde m}_1$ is the key factor that significantly affects the evolution of Eq.~(\ref{eq:diff1}) and (\ref{eq:diff2}) \cite{Buchmuller:2004nz}. In our anarchy model, apart from the overall units, the mass matrix entries are $O(1)$ numbers, so we expect
\begin{equation}
{\tilde m}_1 = O(1) \cdot \frac{{\cal D}^2}{{\cal M}}\ .
\end{equation}
This is just the light-neutrino mass scale. Because in our simulation, ${\cal M}$ is chosen such that $\Delta m_l^2=2.5\times 10^3 $~eV$^2$, we have
\begin{equation}
  \frac{{\cal D}^2}{{\cal M}} = O(1) \cdot \sqrt{\Delta m_l^2} = O(1)
  \cdot 0.05\text{~eV}.
\end{equation}
Therefore most of the time, our model is in the ``strong washout regime" \cite{Buchmuller:2004nz}, since ${\tilde m}_1$ is much larger than the {\it equilibrium neutrino mass} $m_*$:
\begin{equation}
  {{\tilde m}_1} \sim 0.05\text{~eV} \gg {m_*} \sim {10^{ - 3}}\text{~eV}.
    \label{eq:m1condition}
\end{equation}
In this regime, the integral in Eq.(\ref{eq:efficiency}), which is called {\it efficiency factor} $\kappa_f$, should be around \cite{Buchmuller:2004nz}
\begin{equation}
\kappa_f = \int_0^\infty  {dz\frac{D}{{D + S}}\frac{{dN_1}}{{dz}}{e^{
      - \int_z^\infty  {W(z')dz'} }}}  \sim {10^{ - 2}}\ .
\end{equation}
Thus our baryon asymmetry is
\begin{equation}
{\eta_{B0}} = 0.013N_{B - L}^0 \sim 1.3 \times {10^{ - 4}} \cdot
\varepsilon . \label{eq:BaryonAsymmetry}
\end{equation}
To estimate the CP asymmetry $\varepsilon$, we notice that (following the notation of \cite{Buchmuller:2004nz})
\begin{equation}
K \equiv {h^\dag }h \sim {(\frac{{\sqrt 2 }}{v}{\cal D})^2}\left( {\begin{array}{*{20}{c}}
   {{\epsilon ^4}} & {{\epsilon ^3}} & {{\epsilon ^2}}  \\
   {{\epsilon ^3}} & {{\epsilon ^2}} & \epsilon   \\
   {{\epsilon ^2}} & \epsilon  & 1  \\
\end{array}} \right),
\end{equation}
and thus
\begin{eqnarray}
 \varepsilon &=& {\varepsilon _V} + {\varepsilon _M} \sim \frac{3}{{16\pi }}\sum\limits_{k = 1}^3 {\frac{{{\mathop{\rm Im}\nolimits} ({K_{1k}}^2)}}{{{K_{11}}}}} \frac{{{M_1}}}{{{M_k}}} \nonumber \\
 &=& \frac{3}{{16\pi }}\left[ {\frac{{{\mathop{\rm Im}\nolimits} ({K_{12}}^2)}}{{{K_{11}}}}\frac{{{M_1}}}{{{M_2}}} + \frac{{{\mathop{\rm Im}\nolimits} ({K_{13}}^2)}}{{{K_{11}}}}\frac{{{M_1}}}{{{M_3}}}} \right] \nonumber \\
 &\sim& \frac{3}{{16\pi }}
 {\left(\frac{{\sqrt 2 }}{v}{\cal D}\right)^2}
 \left(\frac{{{\epsilon ^6}}}{{{\epsilon ^4}}}{\epsilon ^2} + \frac{{{\epsilon ^4}}}{{{\epsilon ^4}}}{\epsilon ^4}\right) \nonumber \\
 &\sim& \frac{3}{{4\pi }}{\left(\frac{{{\cal D}}}{v}\right)^2}{\epsilon ^4} \sim 3 \times {10^{ - 7}}.
\end{eqnarray}
So the baryon asymmetry is expected to be around ${\eta _{B0}} \sim 1.3 \times {10^{ - 4}} \cdot \varepsilon  \sim 4 \times {10^{ - 11}}$, multiplied by some $O(1)$ factor. This is what we see from Fig.~\ref{fig:LogetaB0}.

Our use of $U(1)$ flavor symmetry breaking plays an essential role to produce the correct order of $\varepsilon$ ($\varepsilon\sim\epsilon^4$) and thus $\eta_{B0}$. It is thus interesting to investigate what would happen if we had a different $U(1)$ charge assignment. An arbitrary charge assignment would be, upon symmetry breaking, equivalent to an arbitrary choice of $D_\epsilon$ parameterized as
\begin{equation}
D_\epsilon = \left( {\begin{array}{*{20}{c}}
   {{\epsilon_1}} & 0 & 0  \\
   0 & {{\epsilon_2}} & 0  \\
   0 & 0 & {{\epsilon_3}}  \\
\end{array}} \right) = {\epsilon_3}\left( {\begin{array}{*{20}{c}}
   {{\epsilon_{r1}}{\epsilon_{r2}}} & 0 & 0  \\
   0 & {{\epsilon_{r2}}} & 0  \\
   0 & 0 & 1  \\
\end{array}} \right),
\end{equation}
where ${\epsilon _1},{\epsilon _2},{\epsilon _3} \lesssim 1$, and we have defined ${\epsilon_{r2}} \equiv {\epsilon _2}/{\epsilon _3}$ and ${\epsilon_{r1}} \equiv {\epsilon _1}/{\epsilon _2}$ for convenience. To make the simplest scenario of leptogenesis work, we need the hierarchy among the heavy-neutrino masses. So we restrict ourselves to the case ${\epsilon_{r1}},{\epsilon_{r2}} \ll 1$.

The Majorana mass matrix now becomes
\begin{eqnarray}
 {m_R} &=& {D_\epsilon }{m_{R0}}{D_\epsilon } \sim \epsilon _3^2\left( {\begin{array}{*{20}{c}}
   {\epsilon_{r1}^2\epsilon_{r2}^2} & {{\epsilon_{r1}}\epsilon_{r2}^2} & {{\epsilon_{r1}}{\epsilon_{r2}}} \\
   {{\epsilon_{r1}}\epsilon_{r2}^2} & {\epsilon_{r2}^2} & {{\epsilon_{r2}}}  \\
   {{\epsilon_{r1}}{\epsilon_{r2}}} & {{\epsilon_{r2}}} & 1  \\
\end{array}} \right) \nonumber \\
 &\sim& \epsilon _3^2\left( {\begin{array}{*{20}{c}}
   1 & {{\epsilon_{r1}}} & {{\epsilon_{r1}}{\epsilon_{r2}}}  \\
   {{\epsilon_{r1}}} & 1 & {{\epsilon_{r2}}}  \\
   {{\epsilon_{r1}}{\epsilon_{r2}}} & {{\epsilon_{r2}}} & 1  \\
\end{array}} \right) \nonumber \\
&\times& \left( {\begin{array}{*{20}{c}}
   {\epsilon_{r1}^2\epsilon_{r2}^2} & 0 & 0  \\
   0 & {\epsilon_{r2}^2} & 0  \\
   0 & 0 & 1  \\
\end{array}} \right) \left( {\begin{array}{*{20}{c}}
   1 & {{\epsilon_{r1}}} & {{\epsilon_{r1}}{\epsilon_{r2}}}  \\
   {{\epsilon_{r1}}} & 1 & {{\epsilon_{r2}}}  \\
   {{\epsilon_{r1}}{\epsilon_{r2}}} & {{\epsilon_{r2}}} & 1  \\
\end{array}} \right),
\end{eqnarray}
which gives
\begin{eqnarray}
M &\sim& \epsilon _3^2\left( {\begin{array}{*{20}{c}}
   {\epsilon_{r1}^2\epsilon_{r2}^2} & 0 & 0  \\
   0 & {\epsilon_{r2}^2} & 0  \\
   0 & 0 & 1  \\
\end{array}} \right), \\
{U_N} &\sim& \left( {\begin{array}{*{20}{c}}
   1 & {{\epsilon_{r1}}} & {{\epsilon_{r1}}{\epsilon_{r2}}}  \\
   {{\epsilon_{r1}}} & 1 & {{\epsilon_{r2}}}  \\
   {{\epsilon_{r1}}{\epsilon_{r2}}} & {{\epsilon_{r2}}} & 1  \\
\end{array}} \right).
\end{eqnarray}
The Dirac mass matrix becomes
\begin{equation}
m_D = m_{D0} D_\epsilon,
\end{equation}
which gives the Yukawa coupling $h$ and $K \equiv h^\dag h$ as
\begin{eqnarray}
 h &=& \frac{{\sqrt 2 }}{v}{m_D}{U_N}^* \nonumber \\
 &\sim& \frac{{\sqrt 2 }}{v}{m_{D0}}{\epsilon _3}\left( {\begin{array}{*{20}{c}}
   {{\epsilon_{r1}}{\epsilon_{r2}}} & {\epsilon_{r1}^2{\epsilon_{r2}}} & {\epsilon_{r1}^2\epsilon_{r2}^2}  \\
   {{\epsilon_{r1}}{\epsilon_{r2}}} & {{\epsilon_{r2}}} & {\epsilon_{r2}^2}  \\
   {{\epsilon_{r1}}{\epsilon_{r2}}} & {{\epsilon_{r2}}} & 1  \\
\end{array}} \right) \\
K &\sim& {(\frac{{\sqrt 2 }}{v}{\cal D})^2}\epsilon _3^2\left( {\begin{array}{*{20}{c}}
   {\epsilon_{r1}^2\epsilon_{r2}^2} & {{\epsilon_{r1}}\epsilon_{r2}^2} & {{\epsilon_{r1}}{\epsilon_{r2}}}  \\
   {{\epsilon_{r1}}\epsilon_{r2}^2} & {\epsilon_{r2}^2} & {{\epsilon_{r2}}}  \\
   {{\epsilon_{r1}}{\epsilon_{r2}}} & {{\epsilon_{r2}}} & 1  \\
\end{array}} \right).
\end{eqnarray}
So our $\varepsilon$ is given by
\begin{eqnarray}
 \varepsilon &=& {\varepsilon _V} + {\varepsilon _M} \sim \frac{3}{{16\pi }}\sum\limits_{k = 1}^3 {\frac{{{\mathop{\rm Im}\nolimits} ({K_{1k}}^2)}}{{{K_{11}}}}} \frac{{{M_1}}}{{{M_k}}} \nonumber \\
 &\sim& \frac{3}{{8\pi }}{(\frac{{\sqrt 2 }}{v}{\cal D})^2}\epsilon _3^2\epsilon_{r1}^2\epsilon_{r2}^2 \nonumber \\
 &\sim& \frac{3}{{4\pi }}{(\frac{{{\cal D}}}{v})^2}\epsilon_1^2.
\end{eqnarray}
We see that under the condition ${\epsilon_{r1}},{\epsilon_{r2}} \ll 1$, $\varepsilon$ is only sensitive to the value of $\epsilon_1$.

On the other hand, the value of ${\tilde m}_1$ is not affected by changing $U(1)$ flavor charge assignments:
\begin{eqnarray}
 {{\tilde m}_1} &\equiv& \frac{{{{({m_D}^\dag {m_D})}_{11}}}}{{{M_1}}} \sim \frac{{{{({m_{D0}}^\dag {m_{D0}})}_{11}}}}{{{(M_{1})_{0}}}}\frac{{\epsilon _1^2}}{{\epsilon _1^2}} \nonumber \\
 &=& \frac{{{{({m_{D0}}^\dag {m_{D0}})}_{11}}}}{{{(M_{1})_{0}}}}={({\tilde m}_{1})_{0}}.
\end{eqnarray}
Here a subscript ``$0$" is used to denote the value when there is no $U(1)$ flavor charge assignment, as we did in Eq.~(\ref{eq:newLagrangian}). So the strong washout condition (Eq.~(\ref{eq:m1condition})) still holds, and we are again led to Eq.~(\ref{eq:BaryonAsymmetry}). Therefore, the baryon asymmetry $\eta_{B0}$ can only be affected through $\varepsilon$, which in turn is only sensitive to $\epsilon_1$, under the condition ${\epsilon_{r1}},{\epsilon_{r2}} \ll 1$.

\section{Correlations between $\eta_{B0}$ and light-neutrino parameters \label{sec:Correlation}}

As we can see from Fig.~\ref{fig:LogetaB0}, the baryon asymmetry is slightly enhanced after applying the Mixing-Split cuts Eq.~(\ref{eq:cut23})-(\ref{eq:cutR}). This indicates some correlation between $\eta_{B0}$ and light-neutrino parameters. To understand this better, we would like to systematically investigate the correlations between $\eta_{B0}$ and the light-neutrino mass matrix $m_\nu=U_\nu m U_\nu^T$. \footnote{The correlation between leptogenesis and light-neutrino parameters has been recently studied in \cite{Jeong:2012zj}. They have some overlap with our results.}

Although both of $\eta_{B0}$ and $m_\nu$ seem to depend on the random inputs $m_R$ and $m_D$ in a complicated way, it is not hard to see that there should be no correlation between $\eta_{B0}$ and $U_\nu$ (This was also pointed out in \cite{Jeong:2012zj}). Recall that we parametrize $m_R$ and $m_D$ as in Eq.~(\ref{eq:paramR})-(\ref{eq:paramD}). And due to the decomposition Eq.~(\ref{eq:decom1})-(\ref{eq:decom2}), there are five independent random matrices: $U_1$, $U_2$, $U_R$, $D_0$ and $D_R$. The first thing to observe is that changing $U_1$ with the other four matrices fixed will not affect $\eta_{B0}$. This is because:
\begin{enumerate}
  \item The baryon asymmetry $\eta_{B0}$ we have been computing in this paper is the total baryon asymmetry, including all the three generations. So $m_D$ enters the calculation of leptogenesis only through the form of the matrix
\begin{equation}
K \equiv h^\dag h = \left(\frac{\sqrt 2}{v}\right)^2 U_N^T m_D^\dag m_D U_N^*,
\end{equation}
with $m_D={\cal D} \cdot U_1 D_0 U_2^\dag$. Obviously $U_1$ cancels in $K$.
  \item Throughout the simulation, we are also applying a built-in cut $\Delta m^2_l=2.5 \times 10^{-3} eV^2$ by choosing the value of ${\cal M}$ to force it. Due to this cut, $m_D$ can potentially affect $\eta_{B0}$ through the value of ${\cal M}$. However, since the actual relation is
\begin{eqnarray}
m_\nu &=& m_D m_R^{-1} m_D^T \nonumber \\
&=& \frac{{\cal D}^2}{{\cal M}} U_1 D_0 U_2^\dag m_R^{-1} U_2^* D_0 U_1^T \nonumber \\
&=& U_\nu m U_\nu^T, \label{eq:Umap}
\end{eqnarray}
we see that changing $U_1$ would only affect $U_\nu$, not $m$. So no further adjustment of ${\cal M}$ is needed when we change $U_1$.
\end{enumerate}
The second point to observe is that any change in $U_\nu$ can be achieved by a left translation
\begin{equation}
U_{\nu a} \to U_{\nu b} = (U_{\nu b} U_{\nu a}^{-1}) U_{\nu a} \equiv U_L U_{\nu a}
\end{equation}
This in turn, can be accounted for by just a left translation in
$U_1$: ${U_{1a}} \to {U_{1b}} = {U_L}{U_{1a}}$, with the other four
random matrices unchanged (see Eq.~(\ref{eq:Umap})). This left
translation in $U_1$ is thus a one-to-one mapping between the
sub-sample generating $U_{\nu a}$ and the sub-sample generating
${U_{\nu b}} = {U_L}{U_{\nu a}}$. Any two events connected through
this one-to-one mapping generate the same value of $\eta_{B0}$,
because changing $U_1$ does not change $\eta_{B0}$. In addition, the
two events have the same chance to appear, because the measure over
$U_1$ is the Haar measure, which is invariant under the left
translation. Thus the sub-sample with $U_\nu=U_{\nu a}$ and
$U_\nu=U_{\nu b}$, for any arbitrary $U_{\nu a}$ and $U_{\nu b}$, will
give the same distribution of $\eta_{B0}$, namely that $\eta_{B0}$ is
independent of $U_\nu$. So immediately we conclude that $\eta_{B0}$
cannot be correlated with the three mixing angles
$\theta_{12},\theta_{23},\theta_{13}$, the CP phase $\delta_{CP}$, or
the phases $\chi_1,\chi_2$.

Three of the Mixing-Split cuts applied to $\eta_{B0}$ are cuts on
mixing angles which we just showed not correlated with $\eta_{B0}$. So
clearly, the enhancement of $\eta_{B0}$ is due to its non-zero
correlation with $R$. To study more detail about the correlation
between $\eta_{B0}$ and the light-neutrino masses $m$, we apply a
$\chi^2$ test of independence numerically to the joint distribution
between $\eta_{B0}$ and quantities related to $m$, including
$\text{lg} R$, $m_{\text{eff}}$ and $m_{\text{total}}$. For each quantity with $\eta_{B0}$, we construct a discrete joint distribution by counting the number of occurrences $O_{ij}$ ($i,j=1,...,10$) in an appropriate $10 \times 10$ partitioning grid. Then we obtain the expected number of occurrences $E_{ij}$ as
\begin{equation}
{E_{ij}} = \frac{1}{n}\left(\sum\limits_{c = 1}^{10} {{O_{ic}}}
\right)
\left(\sum\limits_{r = 1}^{10} {{O_{rj}}}\right),
\end{equation}
where $n$ is the total number of occurrences in all $10 \times 10$ partitions. If the two random variables in question were independent of each other, we would have the test statistic
\begin{equation}
X = \sum\limits_{i,j = 1}^{10} {\frac{{{{({O_{ij}} -
          {E_{ij}})}^2}}}{{{E_{ij}}}}} \ ,
\end{equation}
satisfying the $\chi^2$ distribution with degrees of freedom $(10-1)\times(10-1)=81$. We then compute the probability $P(\chi^2>X)$ for the hypothesis distribution $\chi^2(81)$ to see if the independence hypothesis is likely. Our results from a $n=3,000,000$ sample Monte Carlo are shown in Table.~\ref{tbl:chi2test}.

\renewcommand\arraystretch{1.4}
\begin{table}
\centering
\begin{tabular}{|c|c|c|}\hline
                     & $X$ & $P(\chi^2>X)$ \\ \hline
  $\text{lg} R$       & $\hspace{0.2cm}$ $1.43 \times 10^5$ $\hspace{0.2cm}$ & $\hspace{0.2cm}$ $3.81 \times 10^{-30908}$ $\hspace{0.2cm}$ \\
  $m_{\text{eff}}$   & $8.06 \times 10^3$ & $1.24 \times 10^{-1655\,\,\,}$ \\
  $\hspace{0.2cm}$ $m_\text{total}$ $\hspace{0.2cm}$ & $1.04 \times 10^5$ & $7.24 \times 10^{-22445}$ \\ \hline
\end{tabular}
\caption{\label{tbl:chi2test} $\chi^2$ test of independence between
  $\eta_{B0}$ and $\text{lg} R$, $m_{\text{eff}}$, $m_{\text{total}}$.}
\end{table}
\renewcommand\arraystretch{1}

\begin{figure*}[tpb]
 \centering
 \includegraphics[width=17.5cm]{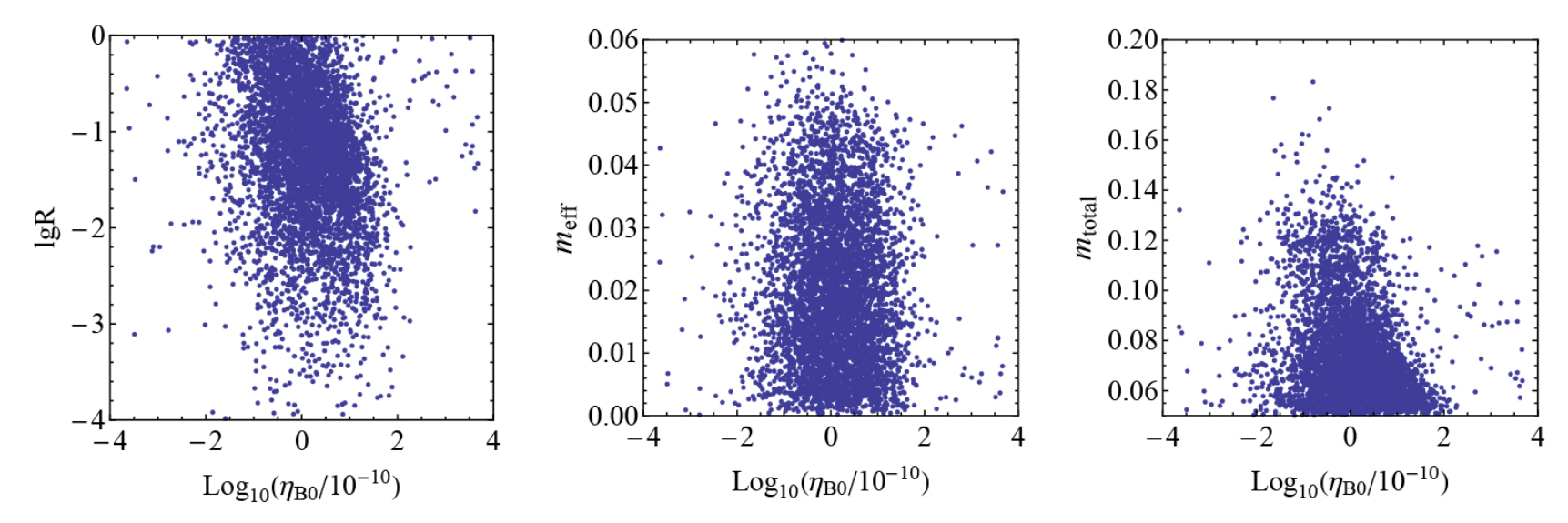}
 \caption{Scatter plots for $\eta_{B0}$ with $\text{lg} R$, $m_\text{eff}$, and $m_\text{total}$. Each plot shows a sample of $5,000$ occurrences. \label{fig:Scatter}}
\end{figure*}

Unambiguously, $\eta_{B0}$ has nonzero correlations with $\text{lg} R$,
$m_\text{eff}$, and $m_\text{total}$. To see the tendency of the
correlations, we draw scatter plots with $5,000$ occurrences
(Fig.~\ref{fig:Scatter}). The plots show that all the three quantities
are negatively correlated with $\eta_{B0}$. For example the left panel
of Fig.~\ref{fig:Scatter} tells us that a smaller $\text{lg} R$ would
favor a larger $\eta_{B0}$. This explains the slight enhancement of
$\eta_{B0}$ after applying Mixing-Split cuts. But as the scatter plots
show, the correlations are rather weak.

\section{Possible Consequences of a Large $m_\text{total}$ \label{sec:MtotalCut}}

As mentioned previously, a recent BOSS analysis suggests
$m_\text{total}$ possibly quite large, $m_\text{total} = 0.36 \pm 0.10
$~eV \cite{Beutler:2014yhv}. Currently their uncertainty is still
large, and thus no conclusive argument can be made. If in future
the uncertainty pins down near its current central value, anarchy
prediction (Fig.~\ref{fig:mtotal}) would be obviously inconsistent
with it and becomes ruled out. On the other hand, if the central value
also comes down significantly, it could still be well consistent with
anarchy prediction.

Without the knowledge of future data, we would like to answer the following question: Assuming the future data be consistent with anarchy, could a relatively large $m_\text{total}$ dramatically change anarchy's predictions on other quantities? For this purpose, we introduce a heuristic $m_\text{total}$ cut:
\begin{equation}
m_\text{total} > 0.1 \text{~eV}, \label{eq:cutMtotal}
\end{equation}
just to get a sense of how much our predictions could be changed if there turns out to be a large but still consistent $m_\text{total}$.

\begin{figure*}[tpb]
 \centering
 \includegraphics[width=13cm]{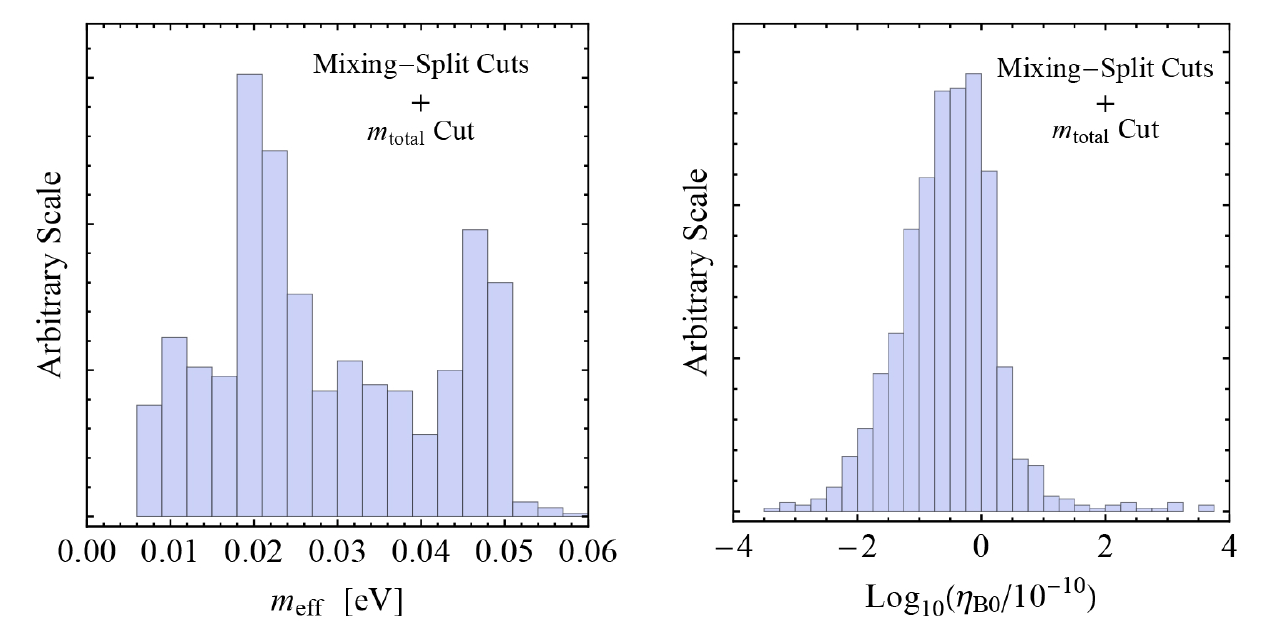}
 \caption{Histograms of $m_\text{eff}$ (left) and $\eta_{B0}$ (right) with $10^4$ occurrences passed both Mixing-Split cuts and the $m_\text{total}$ cut. \label{fig:MtotalCut}}
\end{figure*}

We collect $10^4$ occurrences that pass both the Mixing-Split cuts and the $m_\text{total}$ cut. It turns out that the predictions change quite significantly. We see from Fig.~\ref{fig:hierarchy} that the mass hierarchy prediction is overturned, with normal hierarchy only $40\%$ and inverted scenario more likely. This can be expected from Fig.~\ref{fig:mtotalNI}. The predictions of $m_\text{eff}$ and $\eta_{B0}$ are shown in Fig.~\ref{fig:MtotalCut}. We see that $m_\text{eff}$ exhibits a very interesting bipolar distribution. Its overall expectation value also becomes about an order of magnitude larger than before and thus much less challenging to the neutrinoless double beta decay experiments. The prediction on $\eta_{B0}$ drops by about an order of magnitude, but the observed baryon asymmetry is still very likely to be achieved.

\section{Conclusions \label{sec:Conclusion}}

We have shown that basis independence and free entry independence lead uniquely to Gaussian measure for $m_R$ and $m_D$. We also showed that an approximate $U(1)$ flavor symmetry can make leptogenesis feasible for neutrino anarchy. Combining the two, we find anarchy model successfully generate the observed amount of baryon asymmetry. Same sampling model is used to study other quantities related to neutrino masses. We found the chance of normal mass hierarchy is as high as $99.9\%$. The effective mass of neutrinoless double beta decay $m_\text{eff}$ would probably be well beyond the current experimental sensitivity. The neutrino total mass $m_{\text{total}}$ is a little more optimistic. Correlations between baryon asymmetry and light-neutrino quantities were also investigated. We found $\eta_{B0}$ not correlated with light-neutrino mixings or phases, but weakly correlated with $R$, $m_\text{eff}$, and $m_\text{total}$, all with negative correlation. Possible implications of recent BOSS analysis result have been discussed.

\begin{acknowledgments}
  This work was supported by the U.S. DOE under Contract
  DE-AC02-05CH11231, and by the NSF under grants PHY-1002399 and
  PHY-1316783.  HM was also supported by the JSPS Grant-in-Aid for
  Scientific Research (C) (No.~26400241), Scientific Research on
  Innovative Areas (No.~26105507), and by WPI, MEXT, Japan.
\end{acknowledgments}

\appendix*
\section{Basis Independence and Free Entry Independence Uniquely Lead us to Gaussian Measure}

Let us abstractly write all choices of measure in the form
\begin{equation}
dm = \left( \prod\limits_{ij} dm_{ij} \right) \cdot e^{-f(\left\{
    m_{ij} \right\})}\ , \label{eq:measure}
\end{equation}
where $m$ stands for $m_R$ or $m_D$, $\prod\limits_{ij}$ and $\left\{ m_{ij} \right\}$ run over all the free entries of $m$. We want the form of $f(\left\{ m_{ij} \right\})$ so that the measure above has both basis independence and independence among $m_{ij}$.

Let us first consider $m_D$. For $N \times N$ $m_D$, there are $N^2$ free entries: $m_{11}, m_{12},..., m_{NN}$. For convenience, let us rename them as $x_1, x_2,..., x_n$, where $n=N^2$. Then $\left\{x_i\right\}$ forms an irreducible unitary representation of the basis transformation group $U(3)_L \times U(3)_R$ in flavor space:
\begin{eqnarray}
m_D &\to& m_D' = U_L m_D U_R^\dag , \label{eq:mDtrans} \\
x &\to& x' = \Lambda x , \label{eq:xtrans}
\end{eqnarray}
where $\Lambda = U_L \otimes U_R^*$ is obviously unitary.

Independence of $\left\{m_{ij}\right\}$, namely $\left\{x_i\right\}$, requires $f(\left\{ m_{ij} \right\})$ having the form
\begin{equation}
f(\left\{ x_i \right\}) = f_1(x_1) + f_2(x_2) + \cdots + f_n(x_n).
\end{equation}
Here since $x_i$ are complex arguments, all the functions $f_i(x_i)$ are actually abbreviations of $f_i(x_i,x_i^*)$. Under the transformation of Eq.(\ref{eq:xtrans}), basis independence requires
\begin{equation}
f_1(x_1') + \cdots + f_n(x_n') = f_1(x_1) + \cdots + f_n(x_n). \label{eq:basisIn}
\end{equation}
Taking a derivative with respect to $x_i^*$ yields
\begin{equation}
\sum\limits_j \Lambda_{ji}^*\frac{\partial f_j(x_j')}{\partial
  {x_j'}^*} = \frac{\partial f_i(x_i)}{\partial x_i^*}\ .
\end{equation}
Since $\Lambda$ is unitary, this is the same as
\begin{equation}
\frac{\partial f_i(x_i')}{\partial {x_i'}^*} = \sum\limits_j
\Lambda_{ij}\frac{\partial f_j(x_j)}{\partial x_j^*}\ .
\end{equation}
Thus $\frac{\partial f_i(x_i)}{\partial x_i^*}$ transform in the same way as $x_i$. Because $x_i$ forms an irreducible representation of the transformation Eq.(\ref{eq:xtrans}), the only possibility for $\frac{\partial f_i(x_i)}{\partial x_i^*}$ is
\begin{equation}
\left( {\begin{array}{*{20}{c}}
   {\frac{{\partial {f_1}({x_1})}}{{\partial x_1^*}}}  \\
    \vdots   \\
   {\frac{{\partial {f_n}({x_n})}}{{\partial x_n^*}}}  \\
\end{array}} \right) = {c_a}\left( {\begin{array}{*{20}{c}}
   {{x_1}}  \\
    \vdots   \\
   {{x_n}}  \\
\end{array}} \right),  \label{eq:rep1}
\end{equation}
with $c_a$ an arbitrary constant. Similarly, taking a derivative of Eq.(\ref{eq:basisIn}) with respect to $x_i$ will give us
\begin{equation}
\left( {\begin{array}{*{20}{c}}
   {\frac{{\partial {f_1}({x_1})}}{{\partial {x_1}}}}  \\
    \vdots   \\
   {\frac{{\partial {f_n}({x_n})}}{{\partial {x_n}}}}  \\
\end{array}} \right) = {c_b}\left( {\begin{array}{*{20}{c}}
   {x_1^*}  \\
    \vdots   \\
   {x_n^*}  \\
\end{array}} \right) . \label{eq:rep2}
\end{equation}
Combining Eq.(\ref{eq:rep1}) and (\ref{eq:rep2}) we get
\begin{eqnarray}
f(\left\{ m_{D,ij} \right\}) &=& c_1 (x_1 x_1^* + \cdots + x_n x_n^*) + c_2 \nonumber \\
&=& c_1 \left( \sum\limits_{ij} {\left| m_{D,ij} \right|}^2 \right) + c_2.
 \label{eq:fmD}
\end{eqnarray}

An important condition in this proof is that $x$ forms an irreducible unitary representation of the basis transformation group, $U(3)_L \times U(3)_R$ in the case of $m_D$. For the case of $m_R$, this condition still holds. The relevant basis transformation group for $m_R$ is just $U(3)_R$
\begin{eqnarray}
m_R &\to& m_R' = U_R m_R U_R^T , \\
x &\to& x' = \Lambda x .
\end{eqnarray}
$\Lambda=U_R \otimes U_R$ is reducible in general: $3 \otimes 3 = 6 \oplus 3$, but our $m_R$ is symmetric by definition, which only forms the irreducible subspace ``6" (Note that if $m_R$ were real symmetric, this symmetric subspace ``6" would be further reducible.). So same as in Eq.~(\ref{eq:fmD}), we get
\begin{eqnarray}
&& f(\left\{ {{m_{R,ij}}} \right\}) = c_1 (x_1 x_1^* + \cdots + x_n x_n^*) + c_2 \nonumber \\
&& =c_1 \left( \sum\limits_i {\left| m_{R,ii} \right|}^2 +
  2\sum\limits_{i<j} {\left| m_{R,ij} \right|}^2 \right) + c_2 . \label{eq:fmR}
\end{eqnarray}

In Eq.~(\ref{eq:fmD}) and (\ref{eq:fmR}), $c_1$ corresponds to the freedom of adjusting ${\cal D}$ and ${\cal M}$, while $c_2$ is just an overall normalization factor. Plugging them back into Eq.~(\ref{eq:measure}), we get the Gaussian measure of $m_D$ and $m_R$ as in Eq.~(\ref{eq:mDgaussian}) and (\ref{eq:mRgaussian}).

\bibliography{anarchy}
\bibliographystyle{apsrev4-1}

\end{document}